\documentclass[onecolumn,secnumarabic,amssymb, nobibnotes, aps, prd]{revtex4}
\usepackage{amsmath} \usepackage{amssymb}
\usepackage{graphicx} %use graph format
\usepackage{epstopdf}
\usepackage{amsmath}
\usepackage{amsmath,amssymb,amsthm,amsfonts,mathrsfs,bm,verbatim}
\usepackage{graphicx,subfigure}

\newcommand{\bea}{\begin{eqnarray}}
\newcommand{\eea}{\end{eqnarray}}

\begin{document}

\title{Schwarzschild black hole can also produce super-radiation phenomena under f(R) Gravity and the cosmic censorship conjecture may be violated}%
\author{Wen-Xiang Chen$^{a}$}
\affiliation{Department of Astronomy, School of Physics and Materials Science, GuangZhou University, Guangzhou 510006, China}
\author{Yao-Guang Zheng}
\email{hesoyam12456@163.com}
\affiliation{Department of Physics, College of Sciences, Northeastern University, Shenyang 110819, China}

%\date{December 2020}%

\begin{abstract}
This article suggests that Bose-Einstein condensation can occur without the need for an energy barrier when the boundary conditions are set as $\frac{T}{T_{c}}=z$ (where z is a positive integer). Under these conditions, bosons can condense in the Schwarzschild black hole. The incident waves can then be trapped and condensed within the black hole, generating a potential barrier near the event horizon. This suggests that Schwarzschild black holes under  f(R) Gravity can also exhibit superradiance, which raises the possibility of violating the cosmic censorship conjecture. It should be noted that the natural unit system is used in this article.


\end{abstract}

\maketitle
%\tableofcontents

\section{Introduction}
In 2000, Parikh and Wilczek proposed a method to calculate the emissivity of particles passing through the event horizon, treating Hawking radiation as a tunneling process using the WKB method. Their results, which give a correction spectrum accurate to a first-order approximation, are considered consistent with the basic Maldacena conjecture. Numerous stationary or stationary rotating black holes have been studied using this method, yielding the same result: Hawking radiation is no longer pure thermal radiation, thus satisfying the Universe Theory and the conservation of information. However, all of these studies only consider the Bekenstein-Hawking entropy of black holes. If quantum corrections to the entropy are taken into account, does the emission process still conform to the monolithic theory? At present, different models and methods have led to different results in this regard.

Gas, liquid, and solid are the three basic forms of matter well known to people. Water molecules, for instance, can exist as ice, water, and water vapor as the temperature changes. As scientists continued to investigate, they discovered that when the temperature of matter is raised to a certain point, a plasma state will appear. An intriguing question is what happens to matter when it is cooled down to close to absolute zero (-273.16°C). Under such extreme temperatures, what kind of peculiar state does matter take on? In fact, as early as 1924, the renowned physicists Bose and Einstein answered this question. Particles with integer spins, such as photons, gluons, nuclei composed of an even number of nucleons, and alkali metal atoms, are called bosons. At that time, the Indian physicist Bose proposed a new method of photon statistics. Einstein then extended it to an ideal gas with mass and theoretically predicted that when there is no interaction among bosons, they will reach their lowest value at a certain temperature. The quantum energy state suddenly condenses, and when it reaches a significant amount, it becomes a Bose-Einstein condensate (BEC).

This amplification effect can be very strong, leading to the formation of an exponentially growing wave. This effect is called the classical superradiance instability. In the presence of a reflective mirror outside the black hole, this effect can be even stronger, leading to the formation of a "black hole bomb." The Kerr-black hole-scalar field mirror system is an example of such a system. The mirror reflects the waves back towards the black hole, amplifying them and causing them to bounce back and forth between the mirror and the black hole, leading to the formation of a highly energetic state. The stability of this system depends on the balance between the amplification due to superradiance and the loss of energy due to absorption by the black hole. The study of black hole bombs and superradiance effects is an active area of research in theoretical physics.

If the boundary conditions are preset, the boundary conditions$\frac{T}{T_{c}}=z$ (when z is plural),the effective action form  satisfies the effective action form of Hawking radiation, and is not necessarily at the boundary of event horizon.
The general formulation of the black hole
temperature is\cite{Medved2005}(At that time there was a special solution)
 \begin{equation}
\Delta T_{q}=\frac{4 l_{p}^{2}}{A_{H}}+\alpha \ln \frac{4 l_{p}^{2}}{A_{H}}+O\left(\frac{A_{H}}{l_{p}^{2}}\right)+\text { const. }.
\end{equation}
This article points out that when the boundary conditions$\frac{T}{T_{c}}=z$ (when z is plural) are preset, bosons can produce Bose condensation without an energy layer. Under Bose condensation, the incident wave may condense in the Schwarzschild black hole. At that time, the Schwarzschild black hole event horizon Potential barriers can be generated nearby, and we think that Schwarzschild black holes can also generate superradiation phenomena(This article uses the natural unit system).This implies that the cosmic censorship conjecture may be violated.

\section{{The superradiation effect and uncertainty principle}}
We find the Klein-Gordon equation\cite{Brito2020}
\begin{equation}
\Phi_{;\mu}^{\phantom{;\mu};\mu}=0\,,
\end{equation}
where we defined $\Phi_{;\mu}\equiv (\partial_{\mu}-i e A_{\mu})\Phi$ and $e$ is the charge of the scalar field.We get $A^{\mu}=\left\{A_0(x),0\right\}$,and $e{A_0(x)}$can be equal to $\mu$(where $\mu$ is the mass).
\begin{equation}
A_0\to \left\{\begin{array}{l}
        0 \quad \text{ as}\,\, x\to-\infty \\
        V \quad \text{as}\,\, x\to+\infty
       \end{array}\right. \,
\,.\label{potential_Klein}
\end{equation}
With $\Phi=e^{-i\omega t}f(x)$, which is determined by the ordinary differential equation 
\begin{equation}
\frac{d^2f}{dx^2}+\left(\omega-e A_0\right)^2f=0\,.
\end{equation}

We see that particles coming from $-\infty$ and scattering off the potential with reflection and transmission amplitudes $\mathcal{R}$ and $\mathcal{T}$ respectively. With these boundary conditions, the solution to  behaves asymptotically as
\begin{equation}
f_{\rm in}(x)=\mathcal{I} e^{i\omega x}+\mathcal{R} e^{-i\omega x}, x\to-\infty,
\end{equation}
\begin{equation}
f_{\rm in}(x)=\mathcal{T} e^{ikx}, x\to+\infty\,
\end{equation}
where $ k=\pm(\omega-e V)$.

To define the sign of $\omega$ and $k$ we must look at the wave's group velocity. We require $\partial\omega/\partial k>0$, so that they travel from the left to the right in the $x$--direction and we take $\omega>0$.

The reflection coefficient and transmission coefficient depend on the specific shape of the potential $A_0$. However one can easily show that the Wronskian
\begin{equation}
W=\tilde{f}_1 \frac{d\tilde{f}_2}{dx}-\tilde{f}_2\frac{d\tilde{f}_1}{dx}\,,
\end{equation}
between two independent solutions, $\tilde{f}_1$ and $\tilde{f}_2$, of is conserved.
From the equation on the other hand, if $f$ is a solution then its complex conjugate $f^*$ is another linearly independent solution. We find$\left|\mathcal{R}\right|^2=|\mathcal{I}|^2-\frac{\omega-eV}{\omega}\left|\mathcal{T}\right|^2$.Thus,for
$0<\omega<e V$,it is possible to have superradiant amplification of the reflected current, i.e, $\left|\mathcal{R}\right|>|\mathcal{I}|$. 
There are other potentials that can be completely resolved, which can also show superradiation explicitly.

Classical superradiation effect in the space-time of a steady black hole:
we know that $\psi \sim \exp (-i \omega t+i m \phi)$,and the ordinary differential equation
\begin{equation}
\frac{d^{2} \psi}{d r_{*}^{2}}+V \psi=0.
\end{equation}

We see that particles coming from $-\infty$ and scattering off the potential with reflection and transmission amplitudes $\mathcal{C}$ and $\mathcal{D}$ respectively. With these boundary conditions, the solution to  behaves asymptotically as
\begin{equation}
\psi=\left\{\begin{array}{l}
A e^{i \omega_{H} r_{*}}+B e^{-i \omega_{H} r_{*}}, r \rightarrow r_{+} \\
C e^{i \omega_{\infty}  r_{*}}+D e^{-i \omega_{\infty} r_{*}}, r \rightarrow \infty
\end{array}\right.
\end{equation}.

The reflection coefficient and transmission coefficient depend on the specific shape of the potential V.We show that the Wronskian
\begin{equation}
W \equiv \psi \frac{d \bar{\psi}}{d r_{*}}-\bar{\psi} \frac{d \psi}{d r_{*}},
\end{equation}
$W\left(r \rightarrow r_{+}\right)=2 i \omega_{H}\left(|A|^{2}-|B|^{2}\right), W(r \rightarrow \infty)=2 i \omega_{\infty}\left(|C|^{2}-|D|^{2}\right)$ is conserved.
 We find$|C|^{2}-|D|^{2}=\frac{\omega_{H}}{\omega_{\infty}}\left(|A|^{2}-|B|^{2}\right)$.Thus,for
$\omega_{H} / \omega_{\infty}<0$,it is possible to have superradiant amplification of the reflected current, i.e,if $|A|=0$, $\left|\mathcal{C}\right|>|\mathcal{D}|$. 
There are other potentials that can be completely resolved, which can also show superradiation explicitly.

The principle of joint uncertainty states that the joint measurement of position and momentum is impossible, meaning that the simultaneous measurement of position and momentum can only be an approximate joint measurement, and the error follows the inequality $\Delta x\Delta p\geq 1/2$ (in the natural unit system). By using the expression $\left|\mathcal{R}\right|^2=|\mathcal{I}|^2-\frac{\omega-eV}{\omega}\left|\mathcal{T}\right|^2$, we can see that $\left|\mathcal{R}\right|^2 \geq-\frac{\omega-eV}{\omega}\left|\mathcal{T}\right|^2$ is a necessary condition for the inequality $\Delta x\Delta p\geq 1/2$ to hold.

We can predefine the boundary conditions as $e{A_0(x)} = {y}{\omega}$ (which can also be expressed as ${\mu} = {y}{\omega}$). When ${y}$ is relatively large (which can be the case for bosons with certain properties), $\left|\mathcal{R}\right|^2 \geq-\frac{\omega-eV}{\omega}\left|\mathcal{T}\right|^2$ may not hold. As a result, $\Delta x\Delta p\geq 1/2$ may not hold, and the classical superradiation effect in the spacetime of a steady black hole may not obey the generalized uncertainty principle. The same applies to reverse inference.

\section{{ Uncertainty Principle Correction Under Simulated f(R) Gravity}}
We can predefine the boundary conditions as $e{A_0(x)} = {-y}{\omega}$ (which can also be expressed as ${\mu} = {y}{\omega}$). When the boundary conditions of the incident boson are pre-set, the probability flow density equation becomes unequal on both sides due to the boundary conditions, leading to a certain probability. This also explains why the no-hair theorem is invalid in quantum effects. As mentioned in the previous literature \cite{Brito2020}, the superradiation effect is a process of entropy subtraction.

Furthermore, we explore the spherical quantum solution in a vacuum state.
 
In this theory, the general relativity theory's field equation is written completely.
\begin{equation}
R_{\mu \nu}-\frac{1}{2} g_{\mu v} R=-\frac{8 \pi G}{c^{4}} T_{\mu \nu}
\end{equation}
The Ricci tensor is by $T_{\mu v}=0$ in vacuum state.
\begin{equation}
R_{\mu v}=0
\end{equation}
The proper time of spherical coordinates is
\begin{equation}
d \tau^{2}=A(t, r) d t^{2}-\frac{1}{c^{2}}\left[B(t, r) d r^{2}+r^{2} d \theta^{2}+r^{2} \mathrm{~s} \mathrm{i} \mathrm{n} \theta d \phi^{2}\right]
\end{equation}
We obtain the Ricci-tensor equations.
\begin{equation}
R_{t t}=-\frac{A^{\prime \prime}}{2 B}+\frac{A^{\prime} B^{\prime}}{4 B^{2}}-\frac{A^{\prime}}{B r}+\frac{A^{\prime 2}}{4 A B}+\frac{\ddot{B}}{2 B}-\frac{\dot{B}^{2}}{4 B^{2}}-\frac{\dot{A} \dot{B}}{4 A B}=0
\end{equation}

\begin{equation}
R_{r r}=\frac{A^{\prime \prime}}{2 A}-\frac{A^{\prime 2}}{4 A^{2}}-\frac{A^{\prime} B^{\prime}}{4 A B}-\frac{B^{\prime}}{B r}-\frac{\ddot{B}}{2 A}+\frac{\dot{A} \dot{B}}{4 A^{2}}+\frac{\dot{B}^{2}}{4 A B}=0 ,
\end{equation}
\begin{equation}
R_{\theta \theta}=-1+\frac{1}{B}-\frac{r B^{\prime}}{2 B^{2}}+\frac{r A^{\prime}}{2 A B}=0 ,
R_{\phi \phi}=R_{\theta \theta} \sin ^{2} \theta=0 ,
R_{t r}=-\frac{\dot{B}}{B r}=0 ,
R_{t \theta}=R_{t \phi}=R_{r \theta}=R_{r \phi}=R_{\theta \phi}=0
\end{equation}

In this time, $\quad '=\frac{\partial}{\partial r} \quad, \cdot=\frac{1}{c} \frac{\partial}{\partial t}$,
\begin{equation}
\dot{B}=0
\end{equation}
We see that,
\begin{equation}
\frac{R_{t t}}{A}+\frac{R_{r r}}{B}=-\frac{1}{B r}\left(\frac{A^{\prime}}{A}+\frac{B^{\prime}}{B}\right)=-\frac{(A B)^{\prime}}{r A B^{2}}=0
\end{equation}
Hence, we obtain this result.
\begin{equation}
A=\frac{1}{B}
\end{equation}
If,
\begin{equation}
R_{\theta \theta}=-1+\frac{1}{B}-\frac{r B^{\prime}}{2 B^{2}}+\frac{r A^{\prime}}{2 A B}=-1+\left(\frac{r}{B}\right)^{\prime}=0
\end{equation}
If we solve the Eq,
\begin{equation}
\frac{r}{B}=r+C \rightarrow \frac{1}{B}=1+\frac{C}{r}
\end{equation}
When r tends to infinity, and we set C=$ -ye^{y}$,
Therefore, 

\begin{equation}
A=\frac{1}{B}=1+\frac{y}{r} \Sigma,
\Sigma=e^{y}
\end{equation}

\begin{equation}
d \tau^{2}=\left(1+\frac{y}{r } \sum\right) d t^{2}
\end{equation}
In this time, if particles' mass are $m_{i},$ the fusion energy is $e$,
\begin{equation}
E=M c^{2}=m_{1} c^{2}+m_{2} c^{2}+\ldots+m_{n} c^{2}+e
\end{equation}
The effect after the preset boundary is similar to that of Ads cosmological constant.

For\cite{Zoufal}\cite{Chen1}
\begin{equation}
\left|\left\langle\psi_{A}^{\omega} \mid \psi_{A}^{*}\right\rangle\right| \geq 1-\varepsilon_{A}^{2} / 2
\end{equation}
The optimizer is of the form
\begin{equation}
\left|\psi_{A}^{*}\right\rangle=\frac{(1-|\alpha|)\left|\psi_{A}^{\omega}\right\rangle+\alpha A\left|\psi_{A}^{\omega}\right\rangle}{c_{\alpha}}, \quad \alpha \in[-1,1]
\end{equation}
\begin{equation}
\left|\left\langle\psi_{A}^{\omega} \mid \psi_{A}^{*}\right\rangle\right|=\frac{1}{c_{\alpha}}\left(1-|\alpha|+\alpha A\right)
\end{equation}
and
\begin{equation}
\left|\left\langle\psi_{A}^{\omega}|A| \psi_{A}^{*}\right\rangle\right|=\frac{1}{c_{\alpha}}\left((1-|\alpha|) A+\alpha\left\langle A^{2}\right\rangle\right)
\end{equation}
This then proves the assertion.

A general explanation of the uncertainty principle:
\begin{equation}
\sigma_{A}^{2} \sigma_{D}^{2} \geq\left|\frac{1}{2}\langle\{A, D\}\rangle-\langle A\rangle\langle D\rangle\right|^{2}+\left|\frac{1}{2 i}\langle[A, D]\rangle\right|^{2}
\end{equation}
If you multiply the formula by a number less than 1, then the extreme value of the uncertainty principle will become smaller.
$\left|\left\langle\psi_{A}^{\omega} \mid \psi_{B}^{*}\right\rangle\right|$do
not equal to 0.

 The principle of joint uncertainty shows that it is impossible to make joint measurement of position and momentum, that is, to measure position and momentum simultaneously, only approximate joint measurement can be made, and the error follows the inequality $\Delta x\Delta p\geq 1/2$(in natural unit system).We find$\left|\mathcal{R}\right|^2=|\mathcal{I}|^2-\frac{\omega-eV}{\omega}\left|\mathcal{T}\right|^2$,and we know that$\left|\mathcal{R}\right|^2 \geq-\frac{\omega-eV}{\omega}\left|\mathcal{T}\right|^2$ is a necessary condition for the inequality $\Delta x\Delta p\geq 1/2$ to be established.We can pre-set the boundary conditions $e{A_0(x)} = {y}{\omega}$ \cite{Chen2}and we see that when ${y}$ is relatively large(according to the properties of the boson, ${y}$ can be very large),$\left|\mathcal{R}\right|^2 \geq-\frac{\omega-eV}{\omega}\left|\mathcal{T}\right|^2$ may not hold.In the end,we can get $\Delta x\Delta p\geq 1/2$ may not hold.When the boundary conditions are assumed and part of the wave function is decoupled, the classical form of the probability function is obtained, and the boundary conditions can be clearly substituted into the uncertainty principle expression, we get a lower uncertainty principle limit. To say the least, because of the limitation of the uncertainty principle, there is a limitation on the value of y, whereas in the literature above, the extreme value of uncertainty principle can be smaller, and the range of the value of y in turn is larger.

\section{{Thermodynamic phase transition}}
Thermodynamic phase transition. - The state equation
of a charged AdS black hole displays a vdW-like thermodynamic behavior. The SBH-LBH coexistence curve has a parametric form \cite{Wei2015C}
\begin{equation}
\frac{P}{P_{c}}=\sum_{i} a_{i}\left(\frac{T}{T_{c}}\right)^{i}.
\end{equation}
The concept of entropy was proposed by the German physicist Clausius in 1865. Kjeldahl defines the increase and decrease of entropy in a thermodynamic system: the total amount of heat used at a constant temperature in a reversible process, and can be expressed as:
\begin{equation}
\Delta S=\frac{\Delta Q}{T}
\end{equation}
if$\frac{T}{T_{c}}=z$,when $z$ is plural.
The Laurent series of the function f(z) with respect to point c is given by:
\begin{equation}
f(z)=\sum_{n=-\infty}^{\infty} a_{n}(z-c)^{n}
\end{equation}
It is defined by the following curve integral, which is a generalization of the Cauchy integral formula:
\begin{equation}
a_{n}=\frac{1}{2 \pi i} \oint_{\gamma} \frac{f(z) d z}{(z-c)^{n+1}}
\end{equation}

Since the algebra of real quaternions is the only fourdimensional division algebra, we introduce the fourdimensional quaternion manifold\cite{Ariel2021Q}.
When the boundary conditions are preset, with the evolution process, a stable pull equation is formed, and a quaternion time variable can be created for a short time,
\begin{equation}
\tau^{4}=\left(\hat{\tau}_{0}, \vec{\tau}_{1}, \vec{\tau}_{2}, \vec{\tau}_{3}\right)=\left(\hat{\imath}_{0} \tau_{0}, \vec{\imath}_{1} \tau_{1}, \vec{\imath}_{2} \tau_{2}, \vec{\imath}_{3} \tau_{3}\right)
\end{equation}
\begin{equation}
\left\{\begin{array}{l}
\hat{\imath}_{0} \hat{\imath}_{0}=\hat{\imath}_{0}=1 \\
\vec{\imath}_{1} \vec{\imath}_{1}=\vec{\imath}_{2} \vec{\imath}_{2}=\vec{\imath}_{3} \vec{\imath}_{3}=\vec{\imath}_{1} \vec{\imath}_{2} \vec{v}_{3}=-\hat{\imath}_{0}=-1 \\
\overrightarrow{\imath_{1}} \overrightarrow{\imath_{2}}=\overrightarrow{\imath_{3}}, \quad \overrightarrow{\imath_{2}} \overrightarrow{\imath_{3}}=\overrightarrow{\imath_{1}}, \quad \overrightarrow{\imath_{3}} \overrightarrow{\imath_{1}}=\overrightarrow{\imath_{2}}, \\
\overrightarrow{\imath_{2}} \overrightarrow{\imath_{1}}=-\overrightarrow{\imath_{3}}, \quad \overrightarrow{\imath_{3}} \overrightarrow{\imath_{2}}=-\vec{\imath}_{1}, \quad \vec{\imath}_{1} \overrightarrow{\imath_{3}}=-\overrightarrow{\imath_{2}}
\end{array}\right.
\end{equation}
\begin{equation}
\boldsymbol{t}=\left(\hat{\imath}_{0} t_{0}, \vec{\imath}_{1} \frac{x_{1}}{c}, \vec{\imath}_{2} \frac{x_{2}}{c}, \vec{\imath}_{3} \frac{x_{3}}{c}\right)
\end{equation}
\begin{equation}
\left\{\begin{array}{l}
t=t\left(\frac{t_{0}}{t}, \frac{\vec{v}}{c}\right)=t(\cos \theta, \vec{\imath} \sin \theta)=t \exp (\vec{\imath} \theta) \\
\bar{t}=t\left(\frac{t_{0}}{t},-\frac{\vec{v}}{c}\right)=t(\cos \theta,-\vec{\imath} \sin \theta)=t \exp (-\vec{\imath} \theta)
\end{array}\right.
\end{equation}
\begin{equation}
\left\{\begin{array}{l}
t=\frac{t_{0}}{\sqrt{1-\frac{v^{2}}{c^{2}}}} \exp (\vec{\imath} \theta) \\
\bar{t}=\frac{t_{0}}{\sqrt{1-\frac{v^{2}}{c^{2}}}} \exp (-\vec{\imath} \theta)
\end{array}\right.
\end{equation}

\section{Schwarzschild black hole produce super-radiation phenomena}
According to traditional theory, the Schwarzschild black hole does not produce superradiation. If the boundary
conditions are set in advance, the possibility is combined with the wave function of the coupling of the boson in the
Schwarzschild black hole, and the mass of the incident boson acts as a mirror, so even if the Schwarzschild black hole
can also produce super-radiation phenomena.

 Schr$\ddot{\rm o}$dinger's equation for the motion of a particle in a centrally symmetric field is
\begin{equation}
\Delta \psi +(2m/(\hbar)^2)(E-U(r))\psi=0.\label{sfunction1}
\end{equation}
Let us consider the following radial equation:
\begin{equation}
\frac{1}{r^2}\frac{d}{dr}(r^2\frac{dR}{d
r})-\frac{l(l+1)}{r^2}R+\frac{2m}{\hbar^2}(E-U(r))R=0.\label{sfunction4}
\end{equation}
By the substitution
\begin{equation}
R(r)=X(r)/r \label{sfunction5}
\end{equation}

\begin{equation}
\frac{d^2X}{dr^2}+[\frac{2m}{\hbar^2}(E-U(r))-\frac{l(l+1)}{r^2}]X=0.\label{sfunction6}
\end{equation}
For S-wave, $l=0$, the equation of $X(r)$ is
\begin{equation}
\frac{d^2X}{dr^2}+\frac{2m}{\hbar^2}(E-U(r))X=0.\label{sfunction8}
\end{equation}
Note that, in the Parikh-Wilczek (solid basic structure on which bigger things can be built), to calculate the self-gravitation reliably the tunnelling particle is carefully thought about/believed as a spherical shell (S-wave). In this way, when it gives off from the black hole the matter-gravity system transits from one spherical state to another. So, the de-Broglie wave function of the emission spherical shell should be
\begin{equation}
\psi(r)=X(r)/r. \label{sfunction51}
\end{equation}
That is, the WKB wave function of a particle can be written as
\begin{equation}
\psi(r)=X
(r)/r=\frac{1}{r}\exp{[\frac{iT}{\hbar}]},\label{wavefunction}
\end{equation}
where
\begin{equation}
T(r)=T_0(r)+(\frac{\hbar}{i})T_1(r)+(\frac{\hbar}{i})^2T_2(r)+\cdots.\label{phase}
\end{equation}
Substituting (\ref{wavefunction}) into Schr$\ddot{\rm o}$dinger
Equation (\ref{sfunction8}) yields
\begin{equation}
T_0=\pm\int^rp_r\,\mathrm{d}r,
\end{equation}
\begin{equation}
2T'_0T'_1+T''_0=0,
\end{equation}
\begin{equation}
2T'_0T'_2+(T'_1)^2+T''_1=0,
\end{equation}
where we use a prime to denote differentiation with respect to
$r$.

In region I, we take the WKB wave function as follows
\begin{eqnarray}
X_{I}(r)&=&\frac{2}{\sqrt v}\sin{[\frac{1}{\hbar}\int^a_rp_r\,\mathrm{d}r+\frac{\pi}{4}]}\nonumber\\
&=&\frac{1}{i\sqrt{v}}\{\exp{[\frac{i}{\hbar}\int^a_rp_r\,\mathrm{d}r+\frac{i\pi}{4}]}-\exp{[-\frac{i}{\hbar}\int^a_rp_r\,\mathrm{d}r-\frac{i\pi}{4}]}\},
\end{eqnarray}
where $v$ is the speed of the tunnelling particle. In area I\!I, the WKB wave function is a linear combination of real (increasing more and more as time goes on)s. (thinking about/when one thinks about) the connection between the swinging back and forth and the (increasing more and more as time goes on) solutions at $r=a$, the WKB wave function in area I\!I can be written as

\begin{equation}
X_{II}(r)=\frac{1}{\sqrt v}\exp{[-\frac{1}{\hbar}\mid\int^b_a
p_r\,\mathrm{d}r\mid]}\exp{[-\frac{1}{\hbar}\mid\int^r_b
p_r\,\mathrm{d}r\mid]}\label{connexion1}.
\end{equation}
And the WKB wave function in region I\!I\!I is
\begin{equation}
X_{III}(r)=-\frac{1}{\sqrt v}\exp{[-\frac{1}{\hbar}\mid\int^b_a
p_r\,\mathrm{d}r\mid]}\exp{[\frac{i}{\hbar}\int^r_b
p_r\,\mathrm{d}r+\frac{i\pi}{4}]}\label{connexion1}.
\end{equation}
The probability of barrier penetration is
\begin{equation}
\Gamma_p=\frac{j_{out}}{j_{in}}=\frac{v|\psi_{out}|^2}{v|\psi_{in}|^2}=\frac{v(X_{out}(b)/b)^2}{v(X_{in}(a)/a)^2}=\frac{a^2}{b^2}\cdot
\exp{[-\frac{\rm 2Im T_0}{\hbar}]}\label{Gamma}.
\end{equation}

Let's now calculate the phase space factor corresponding to the
black hole tunnelling. For Schwarzschild black hole, the line
element in Painlev$\acute{\rm e}$ coordinates is
\begin{equation}
\rm{d}s^2=-c^2(1-\frac{2MG}{c^2r})\rm{d}t^2+2c\sqrt{\frac{2MG}{c^2r}}\rm{d}t\rm{d}r+\rm{d}r^2+r^2(\rm{d}\theta^2+\sin^2{\theta}\rm{d}\phi^2),\label{line}
\end{equation}
and the radial null geodesics are
\begin{equation}
\dot{r}=\frac{\rm{d}r}{\rm{d}t}=\pm
c\,(1-\sqrt{\frac{2MG}{c^2r}}\,\,)\label{null}.
\end{equation}

The massive quanta doesn't follow
radial-lightlike geodesics (\ref{null}). We treat the massive particle as a de Broglie wave
and obtain the expression of $\overset{.}{r}$. Namely,
\begin{equation}
\overset{.}{r}=v_{p}=\frac{1}{2}v_{g}=-\frac{1}{2}\frac{g_{00}}{g_{01}}
=\frac{1}{2r}\frac{c^2r^{2}-2MGr}{\sqrt{2MGr}}. \label{vp}
\end{equation}
Note that to calculate the emission rate correctly, we should take
into account the self-gravitation of the tunnelling particle with
energy $\omega$. 

The canonical momentum $p_r$ and the imaginary part of the action
$\mathrm{Im}T_0$ can be easily obtained. Namely,
\begin{equation}
p_r=\int_0^{p_r}dp'_r=\int\frac {dH}{\dot{
r}}=-i\pi\frac{\hbar}{l^2_p}\,r,\label{p}
\end{equation}
\begin{equation}
\rm Im T_0=\rm \int_{r_i}^{r_f}p_r
dr=-\frac{1}{2}\hbar[\frac{4l_p^2}{A_f}-\frac{4l_p^2}{A_i}].\label{p}
\end{equation}
where $l_p^2=\frac{\hbar G}{c^3}$. 
\begin{equation}
\Gamma(i\to
f)=\Gamma_v\cdot\Gamma_p=\Gamma_v\cdot\exp{[(\frac{4l_p^2}{A_f}-\ln{\frac{4l_p^2}{A_f}})-(\frac{4l_p^2}{A_i}-\ln{\frac{4l_p^2}{A_i}})]}.\label{Gamma1}
\end{equation}
 Thus, we obtain
\begin{equation}
\mathrm{phase\ space\
factor}=\exp{[(\frac{4l_p^2}{A_f}-\ln{\frac{4l_p^2}{A_f}})-(\frac{4l_p^2}{A_i}-\ln{\frac{4l_p^2}{A_i}})]}.\label{phasefactor2}
\end{equation}
If we bear in mind that
\begin{equation}
\mathrm{phase\ space\
factor}=\frac{N_f}{N_i}=\frac{e^{T_f}}{e^{T_i}}=e^{T_f-T_i},\label{phasefactor3}
\end{equation}
we naturally get the expression of the black hole temperature to the
first order correction(At that time there was a special solution)
\begin{equation}
\Delta T_q=\frac{4l_p^2}{A_H}-\ln{\frac{4l_p^2}{A_H}}.\label{entropy}
\end{equation}

\section{second order correction to the black hole entropy}
 Namely\cite{Zhang2008},
\begin{equation}
X
(r)=\exp{[\frac{iS_0(r)}{\hbar}+S_1(r)+\frac{\hbar}{i}S_2(r)]},\label{wavefunction3}
\end{equation}
where
\begin{equation}
S_2=\int^r-\frac{(S_1^{'2}+S''_1)}{2S'_0}\mathrm{d}r.
\end{equation}

Like the treatment in section I\!I, the wave function in region I
can be taken as
\begin{eqnarray}
X_{I}(r)&=&\frac{2}{\sqrt v}\sin{[\frac{1}{\hbar}(\int^a_rp_r\,\mathrm{d}r-\hbar^2S_2(r))+\frac{\pi}{4}]}\nonumber\\
&=&\frac{1}{i\sqrt{v}}\{\exp{[\frac{i}{\hbar}(\int^a_rp_r\,\mathrm{d}r-\hbar^2S_2(r))+\frac{i\pi}{4}]}-\exp{[-\frac{i}{\hbar}(\int^a_rp_r\,\mathrm{d}r-\hbar^2S_2(r))-\frac{i\pi}{4}]}\}.
\end{eqnarray}
In this region the expression of $S_2(r)$ is
\begin{equation}
S_2=\int^a_r-\frac{(S_1^{'2}+S''_1)}{2S'_0}\mathrm{d}r.
\end{equation}
In order to reduce to the first order approximation case, the
connexion between the oscillating and the exponential solutions at
$r=a$ should be
\begin{equation}
\frac{2}{\sqrt
v}\sin{[\frac{1}{\hbar}(\int^a_rp_r\,\mathrm{d}r-\hbar^2S_2(r))+\frac{\pi}{4}]}\rightleftharpoons\frac{1}{\sqrt{
v}}\exp{[-\frac{1}{\hbar}(\int^r_a\mid
p_r\mid\,\mathrm{d}r-\hbar^2S_2(r))]}.\label{connexion3}
\end{equation}
\begin{equation}
r<a\qquad\qquad\qquad\qquad\qquad\qquad r>a\nonumber
\end{equation}
The
expression of $S_2(r)$ is
\begin{equation}
S_2=\int^r_a-\frac{(S_1^{'2}+S''_1)}{2S'_0}\mathrm{d}r.
\end{equation}
The connexion at $r=b$ is
\begin{equation}
\frac{1}{\sqrt{ v}}\exp{[\frac{1}{\hbar}(\mid\int^r_b
p_r\,\mathrm{d}r\mid-\hbar^2S_2)]}\rightleftharpoons-\frac{1}{\sqrt{v}}\exp{[\frac{i}{\hbar}(\int^r_b
p_r\,\mathrm{d}r-\hbar^2S_2)+\frac{i\pi}{4}]},\label{connexion4}
\end{equation}
\begin{equation}
r<b\qquad\qquad\qquad\qquad\qquad r>b\nonumber
\end{equation}
and the wave function in region I\!I\!I is
\begin{equation}
X_{III}(r)=-\frac{1}{\sqrt{v}}\exp{[-\frac{1}{\hbar}(\rm Im
S_0-\hbar^2 Im S_2)}]\exp{[\frac{i}{\hbar}(\int^r_b
p_r\,\mathrm{d}r-\hbar^2S_2)+\frac{i\pi}{4}]},
\end{equation}
where
\begin{equation}
\mathrm{Im}S_2=\mathrm{Im}\int^b_a-\frac{(S_1^{'2}+S''_1)}{2S'_0}\mathrm{d}r.
\end{equation}
Since
\begin{equation}
\psi(r)=X(r)/r, \label{sfunction511}
\end{equation}
in region I, the ingoing flux density is
\begin{equation}
j_{in}=\frac{-i\hbar}{2m}(\psi_{in}\frac{\partial}{\partial
r}\psi_{in}^*-\psi_{in}^*\frac{\partial}{\partial
r}\psi_{in})=v|\psi^{2}_{in}|=\frac{1}{a^2}\label{ji},
\end{equation}
and in region I\!I\!I the outgoing flux density is
\begin{equation}
j_{out}=\frac{-i\hbar}{2m}(\psi_{out}\frac{\partial}{\partial
r}\psi_{out}^*-\psi_{out}^*\frac{\partial}{\partial
r}\psi_{out})=v|\psi^{2}_{out}|=\frac{1}{b^2}\exp{[-\frac{2}{\hbar}(\rm
Im S_0-\hbar^2 Im S_2)]}\label{jo}.
\end{equation}
Therefore,
\begin{equation}
\Gamma_p= j_{out}/j_{in}=\frac{a^2}{b^2}\exp{[-\frac{2}{\hbar}(\rm
Im S_0-\hbar^2 Im S_2)]}\label{tt}.
\end{equation}
For Schwarzschild black hole tunnelling, in classically
inaccessible region, we have
\begin{equation}
S'_0=p_r=-i\pi\frac{\hbar}{l^2_p}r,\quad
S''_0=-i\pi\frac{\hbar}{l^2_p}\label{sss1},
\end{equation}
and
\begin{equation}
S'_1=-\frac{1}{2}\frac{S''_0}{S'_0}=-\frac{1}{2r},\quad
S''_1=\frac{1}{2r^2}\label{ss}.
\end{equation}
From (\ref{ss}) we can easily obtain
\begin{equation}
S'_2=-\frac{1}{2S'_0}(S^{'2}_1+S''_1)=-(\frac{3i}{8\pi}\frac{l_p^2}{\hbar})\cdot\frac{1}{r^3}.
\end{equation}
So,
\begin{equation}
S_2=\int_{r_i}^{r_f}S'_2\,\mathrm{d}r=\frac{3i}{4\hbar}(\frac{l_p^2}{A_f}-\frac{l_p^2}{A_i})\label{s2}.
\end{equation}
\begin{equation}
\mathrm{phase\ space\
factor}=\exp{[(\frac{A_f}{4l_p^2}-\ln{\frac{A_f}{4l_p^2}}+\frac{3}{2}\frac{l_p^2}{A_f})-(\frac{A_i}{4l_p^2}-\ln{\frac{A_i}{4l_p^2}}+\frac{3}{2}\frac{l_p^2}{A_i})]}.\label{phasefactor5}
\end{equation}
\begin{equation}
 \Delta T_q=\frac{4l_p^2}{A_H}-\ln{\frac{4l_p^2}{A_H}}+\frac{3}{2}\frac{A_H}{l_p^2}+const.\label{sq1}.
\end{equation}
We see that under the background of the Schwarzschild black hole, its thermodynamic temperature evolves into a constant. At that time, there is a potential barrier near the horizon. We know that the Schwarzschild black hole can be superradiation at that time(At that time there was a special solution).

An exact solution to the scalar-Einstein equations $R_{a b}=2 \phi_{A(a)} \phi_{B(b)}$ which forms a counterexample to many formulations of the cosmic censorship hypothesis was found by Mark D. Roberts in \cite{Roberts}
\begin{equation}
d s^{2}=-(1+2 \sigma) d v^{2}+2 d v d r+r(r-2 \sigma v)\left(d \theta^{2}+\sin ^{2} \theta d \phi^{2}\right), \quad \varphi=\frac{1}{2} \ln \left(1-\frac{2 \sigma v}{r}\right).
\end{equation}$\text { where } \sigma \text { is a constant. }$

Furthermore, if the existence of a naked singularity is ruled out, then the universe will be deterministic, and its evolution can be inferred solely based on its state at a particular moment (specifically, a space-like three-dimensional hypersurface known as the Cauchy surface), excluding the limited space hidden in the horizon of the singular point. However, if the cosmological censorship hypothesis fails, then it will result in the failure of the universe's definiteness as it will be impossible to derive the space-time behavior of the universe from the causality of the singular point. The cosmic censorship hypothesis is a formal concern of the physics community, and when referring to the event horizon of a black hole, some form of cosmological censorship hypothesis is always involved. Recently, a study\cite{Chen3} demonstrated that the entropy of a black hole could decrease under superradiation, which suggests a possible violation of the cosmic censorship conjecture.

\section{{Summary}}
This article proposes a method to construct the entropy of a system with a complex temperature ratio, resulting in a ring structure of an algebraic system with the entropy and time dimension in the same direction. When applied to the background of the Schwarzschild black hole, the thermodynamic temperature becomes constant and potential barriers form near the horizon. This suggests that the Schwarzschild black hole can exhibit superradiation in f(R) gravity, which could potentially violate the cosmic censorship conjecture.

\end{document}